%% file: main.tex
\documentclass[]{spie}  

\usepackage{amsmath,amsfonts,amssymb}
\usepackage{graphicx}
\usepackage[colorlinks=true, allcolors=blue]{hyperref}
	\usepackage[perpage]{footmisc} 
	\usepackage{amsmath}
	\usepackage{mathrsfs} 
   	\usepackage{aas_macros} 
	\usepackage{amssymb}
	\usepackage{amsfonts}
	\usepackage{graphicx}  
	\usepackage{setspace}
	\usepackage{cancel}
    \usepackage[left=0.875in,right=0.875in,top=1in,bottom=1.25in]{geometry}
	\usepackage{enumitem}
	\usepackage{etoolbox}
	\usepackage{siunitx}	
	\usepackage{multirow,bigdelim}
	\usepackage{textcomp}
	\usepackage{float}
	\usepackage{color, colortbl} 
	\usepackage{booktabs} 
	\usepackage{wrapfig} 
	\usepackage{lipsum}
	\usepackage[font=singlespacing]{caption}
	\usepackage{wasysym}
	\usepackage{pdflscape}
	\usepackage{afterpage}
	\usepackage{url}
	\usepackage{upgreek}
	\usepackage{lscape} 
	\usepackage{subcaption} 
	\usepackage{mathtools}          
	
	\usepackage{perpage}
    \MakePerPage{footnote}

	\AtBeginEnvironment{quote}{\singlespacing\small\it}

	\usepackage{xspace}
    \newcommand{\um}{$\upmu$m\xspace}	
	
	\renewcommand{\deg}{$^{\circ}$\xspace}
	\newcommand{\degC}{$^{\circ}$C\xspace}

    \newcommand{\arcsec}{$^{\prime\prime}$}	
    
    \newcommand{\rprocess}{\emph{r}-process\xspace}
    \newcommand{\degsq}{deg$^2$\xspace}
    \newcommand{\textsim}{$\sim$}

	\setlist[enumerate]{label*=\arabic*.}
	\setlist[enumerate,1]{label*=\arabic*.,font=\bfseries,before=\bfseries}
	\setlist[enumerate,2]{label*=\arabic*.,font=\normalfont,before=							\normalfont}
	\setlistdepth{9}

\title{The wide-field infrared transient explorer (WINTER)}

\author[a]{Nathan P. Lourie}
\author[c]{John W. Baker}
\author[c]{Richard S. Burruss}
\author[a]{Mark Egan}
\author[a]{Gábor Fűrész}
\author[a,b]{Danielle Frostig}
\author[a]{Allan A. Garcia-Zych}
\author[c]{Nicolae Ganciu}
\author[a]{Kari Haworth}
\author[a]{Erik Hinrichsen}
\author[c]{Mansi M. Kasliwal}
\author[c]{Viraj R. Karambelkar}
\author[a]{Andrew Malonis}
\author[a,b]{Robert A. Simcoe}
\author[d]{Jeffry Zolkower}

\affil[a]{MIT Kavli Center for Astrophysics and Space Research, Massachusetts Institute of Technology, 77 Massachusetts Ave, Cambridge, MA 02139, USA}
\affil[b]{MIT Department of Physics, 77 Massachusetts Ave., Cambridge, MA 02139, USA}
\affil[c]{Division of Physics, Math, and Astronomy, California Institute of Technology, 1200 E California Blvd, Mail Code 249-17, Pasadena, CA 91125, USA}
\affil[d]{Caltech Optical Observatories, California Institute of Technology, 1200 E California Blvd., Mail Code 11-17, Pasadena, CA 91125, USA}

\authorinfo{Further author information: (Send correspondence to N.P.L)\\N.P.L.: E-mail: nlourie@mit.edu, Telephone: +1 617 324 1194}

\begin{document} 
\maketitle

\begin{abstract}
The Wide-Field Infrared Transient Explorer (WINTER) is a new infrared time-domain survey instrument which will be deployed on a dedicated 1 meter robotic telescope at the Palomar Observatory. WINTER will perform a seeing-limited time domain survey of the infrared (IR) sky, with a particular emphasis on identifying \rprocess material in binary neutron star (BNS) merger remnants detected by LIGO. We describe the scientific goals and survey design of the WINTER instrument. With a dedicated trigger and the ability to map the full LIGO O4 positional error contour in the IR to a distance of 190 Mpc within four hours, WINTER will be a powerful kilonova discovery engine and tool for multi-messenger astrophysics investigations. In addition to follow-up observations of merging binaries, WINTER will facilitate a wide range of time-domain astronomical observations, all the while building up a deep coadded image of the static infrared sky suitable for survey science.
WINTER's custom camera features six commercial large-format Indium Gallium Arsenide (InGaAs) sensors and a tiled optical system which covers a \textgreater1-square-degree field of view with 90\% fill factor. The instrument observes in Y, J and a short-H (Hs) band tuned to the long-wave cutoff of the InGaAs sensors, covering a wavelength range from 0.9 – 1.7 microns. We present the design of the WINTER instrument and current progress towards final integration at the Palomar Observatory and commissioning planned for mid-2021.

\end{abstract}


\input{Introduction}

\input{FocalPlanes}
\input{Camera}

\input{Observatory}

\input{Observing}
\input{Status}
\input{Conclusion}

\acknowledgments 
WINTER's construction is made possible by the National Science Foundation under MRI grant number AST-1828470.  We also acknowledge significant support from the California Institute of Technology, the Caltech Optical Observatories (COO), the Bruno Rossi Fund of the MIT Kavli Institute for Astrophysics and Space Research, and the MIT Department of Physics and School of Science. The collaboration also acknowledges the ongoing support and contributions to the observatory by the COO staff. Eric Bellm from the University of Washington, Reed Riddle from COO, and Javier Romualdez from Princeton University contributed support and guidance on the WINTER controls software.

\bibliography{main} 
\bibliographystyle{spiebib} 

\end{document}

%% file: Introduction.tex
\section{Introduction}
WINTER (The Wide-field Infrared Transient Explorer) is a new robotic infrared time-domain survey instrument which will operate behind a dedicated 1 m aperture telescope at Palomar Observatory. The instrument is currently being assembled and integrated at the MIT Kavli Institute, and will be deployed to Palomar Observatory in mid-2021. In this paper we describe the driving science goals of WINTER and how these goals drive the architecture of the instrument. This paper is published alongside companion papers describing the derivation of the WINTER engineering requirements from scientific goals (Ref.~\citenum{winter_reqs}), the design and testing of the optomechanical structures and lens mounts (Ref.~\citenum{winter_optics}), and the InGaAs sensors and readout electronics (Ref.~\citenum{winter_sensors}).

The design of the WINTER sensor and optical architectures are described in Secs. \ref{sec: detectors} and \ref{sec: camera} respectively. Sec. \ref{sec: observatory} describes the observatory robotic control system, and Sec. \ref{sec: observing} gives an overview of the observation strategy and automated scheduling. Finally, in Sec. \ref{sec: status}, we describe the current status of the instrument and observatory as of November 2020.

\subsection{Science Goals}
Time-domain infrared surveys offer a key window into a host of astrophysical phenomena, especially in the age of multi-messenger astrophysics. Infrared observations are one of the most powerful tools for detecting kilonovae, the thermal emission from rapid neutron capture (\rprocess) nucleosythesis in the ejecta of binary neutron star (BNS) mergers. The LIGO-Virgo experiments' detection of gravitational waves (GW) from the BNS merger GW170817\cite{gw170817} and subsequent detection of electromagnetic (EM) counterparts\cite{kn170817, em170817_Kasliwal_2017} across the spectrum from the x-ray to the radio suggest that these merger events may be the dominant mode of $r$-process element production in the Universe\cite{Kasen_2017}. While GW170817 is the only confirmed observation of a kilonova associated with a BNS merger, improvements in sensitivity for the fourth observing run of LIGO (LIGO O4) are expected to yield 1-2 NS-NS mergers per month.\cite{ligo_observing_prospects_2016} Theoretical models predict that infrared kilonova are ubiquitous in these merger events, and unlike optical emission which is short-lived (\textless1 week) and highly orientation-dependent, the IR signal is long-lived (\textgreater2 weeks) and independent of geometry, opacity, remnant lifetime, and mass ratio.\cite{Kasen_2013,Barnes_2016} 

The primary science goal of WINTER is to undertake a systematic and unbiased search for infrared counterparts to detect lanthanide-rich kilonovae throughout the full LIGO O4 search volume. With three filters operating at the Y, J, and a shortened H (Hs) bands (centered at 1.02, 1.25, and 1.60 \um respectively), WINTER will record weeks-long lightcurves of massive merger events detected by LIGO offering new insight into the mass fraction of lanthanides in the ejecta of these events. This data will help answer fundamental questions such as (a) Are NS-NS mergers the only site of \rprocess nucleosynthesis? (b) Do BNS mergers produce the same relative abundance ratios seen in the solar neighborhood? (c) Are the third-peak \rprocess elements synthesized? (d) What elements are made when a NS merges with a stellar black hole (BH)? WINTER's observing plan (described in more detail in Section \ref{sec: observing}) will prioritize these LIGO detections, allowing follow-up of a larger sample of GW events. With 100\% of observing time available in both the bright and dark lunar cycle, it can follow up all GW events, including those with lower LIGO detection significance. Additionally, WINTER's large (\textsim1 \degsq) field of view (FoV) is well matched to fast mapping of the typical 10-20 \degsq uncertainty contour from a LIGO-Virgo merger alert. WINTER's sensitivity, wide FoV, dedicated telescope, and flexible scheduler make it unmatched in ability to systematically survey kilonova emission in the infrared.

Beyond kilonovae, WINTER's synoptic time-domain infrared survey will provide a new perspective on intrinsically red or obscured astrophysical events. Building on observational and analysis techniques developed for the pathfinder Palomar Gattini-IR (PGIR) survey\cite{Gattini_De_2020}, WINTER will be able to detect and classify transient events such as stellar mergers, tidal disruption events, failed supernovae, and even exoplanet transits around low-mass stars.\cite{kasliwal2019dynamic} WINTER is the first seeing-limited instrument dedicated to systematic infrared time-domain searches and will survey the available Northern sky every two weeks. 

\subsection{Instrument Overview}
Despite the large number of optical transient survey instruments commissioned (or soon to be) over the past decade (e.g. ZTF\cite{ztf_overview_Bellm_2018}, PanSTARRS\cite{PanSTARRS}, ATLAS\cite{ATLAS}, DECam\cite{DECam}, HSC\cite{HSC}, Rubin Observatory/LSST\cite{LSST}), there have been few comparable efforts in the infrared. This is largely due to the high cost (\textsim 30x the cost-per-pixel of CCDs) of state-of-the-art HgCdTe sensors and their associated cryogenic instruments. Fast optical systems can widen the field of view, but are costly as well, and incur an associated reduction in sensitivity because of the high sky background noise in the infrared. 

WINTER achieves background-limited low-cost wide-field imaging  using a custom camera based on an array of newly-developed commercial indium-gallium-arsenide (InGaAs) sensors. While InGaAs sensors have been used in defense applications for years, they have only recently realized the low dark current needed for background-limited astronomical applications, when cooled with a simple thermoelectric cooler (TEC). Using custom readout electronics based on commercially-available field programmable gate array (FPGA) modules, WINTER is able to make multiple non-destructive sensor reads per second, helping to reduce read noise. WINTER's 1.0\deg x 1.2\deg FoV matches that of VIRCAM on the VISTA telescope\cite{vircam_vista}, currently the largest IR focal plane, but with a twofold increase in fill factor and at a \textsim10x reduction in cost. 

\begin{figure}[ht]
\centering
\includegraphics[width=\linewidth]{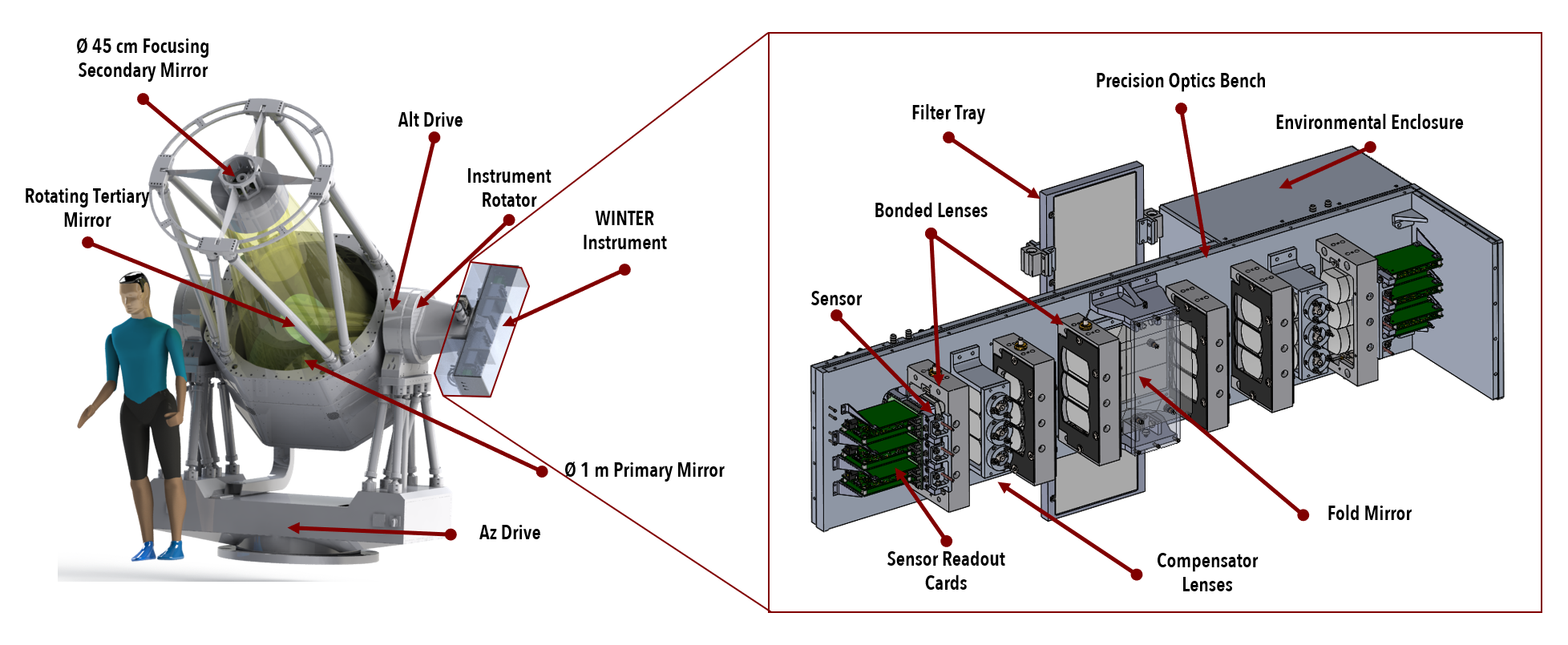}
\caption[WINTER Overview] {CAD rendering showing an overview of the WINTER instrument mounted on its telescope. Inset shows a cutaway view of the instrument, with important subsystems and optical elements labeled.}
\label{figure:winter_overview}
\end{figure}

WINTER couples its split focal plane array to a commercial-off-the-shelf (COTS) 1 m aperture telescope. The WINTER camera's unique optical system slices the telescope focal plane into six identical optical channels, achieving a near-100\% fill factor despite not being able to directly abut the packaged InGaAs focal planes. The camera optics reimage the telescope's F/6 beam down to F/3 at the focal planes to optimally sample the expected seeing at Palomar. The telescope is robotically controlled with custom software which autonomously parses weather data from the site, produces an optimized schedule for each night, executes observing routines and continually logs housekeeping data. With the fast pointing response of the telescope and low instrumental overheads we expect an on-sky duty cycle of \textgreater90\%  and correspondingly high survey efficiency. An overview of the WINTER instrument is shown in Fig. \ref{figure:winter_overview}, with individual subsystems described in detail in the following sections.

%% file: FocalPlanes.tex
\section{Detector Architecture}
\label{sec: detectors}
\subsection{Detectors}
The custom camera for WINTER features six commercial large high-definition (HD) format (1920 x 1080 pixels) AP1020 Indium Gallium Arsenide (InGaAs) sensors currently being developed for our group by FLIR Electro-Optical Systems. These hybridized CMOS focal plane arrays will be the largest and highest-performing InGaAs detectors on the market.  

The system builds on lessons learned from the development of a prototype camera built by the MIT group and tested on the 100-inch DuPont Telescope at Las Campanas Observatory in Chile. This prototype instrument used a single 640 x 512 pixel FLIR AP1121 InGaAs sensor, which is a direct predecessor of the AP1020 sensors being developed for WINTER, with similar component architecture at both the level of the individual CTIA pixel unit cell and the output amplifier. This ``DuPont Prototype" instrument demonstrated sky-background-limited performance in J-band with InGaAs sensors for the first time.\cite{Simcoe_2019} Each of WINTER's sensors is an $8\times$  scaled-up version of the AP1121, with 1920 x 1080 pixels on a 15 \um pitch, and eight output channels. Additional modifications to the sensor package were made to facilitate a closer packing of the sensors in the camera focal plane, including a redesign of the mounting footprint, and the location and orientation of the pinch tube (see Section \ref{subsec: thermal}) from which the sensor package is pumped to vacuum during assembly. 

\begin{figure}[ht]
\centering
\includegraphics[width=0.75\linewidth]{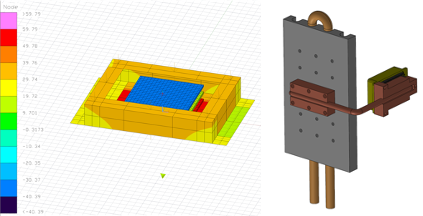}
\caption[Sensor Overview] {\textbf{Left} Thermal model made with ThermalDesktop\footnote{Cullimore and Ring Technologies, Boulder, CO USA} of a TEC in the WINTER sensor packaging, used for calibrating thermal contact resistances and sizing heat pipes. \textbf{Right:} CAD rendering of a prototype heat sinking assembly for the WINTER sensors, showing the sintered copper/water heat pipe soldered to two high purity copper heat sinks bolted to the sensor at one end, and a liquid-cooled heat exchanger on the other.}
\label{figure:sensor}
\end{figure}%

\subsection{Readout}
\label{subsection: readout}
Each of the six sensors interfaces to its own readout system being developed at MIT\cite{winter_sensors}, which features custom amplifier electronics, an FPGA interface, and readout software to implement non-destructive (i.e. sample-up-the-ramp\cite{sutr_glendinning_chapman_1990} (SUTR) or Fowler\cite{fowler_1990}) sampling modes which reduce the effective read noise by using multiple reads to fit the real-time count rate. A Xilinx\footnote{Xilinx Inc., San Jose, CA USA} FPGA was selected in order to build on firmware developed for the DuPont Prototype.\cite{Sullivan_2013} The particular FPGA interface, an OpalKelly\footnote{OpalKelly Inc., Portland, OR USA} ZEM7310-A200 featuring a Xilinx Artix-7 FPGA, balances cost with memory (1 Gb of on-board RAM, and 13.4 Mb of FPGA RAM) in order to allow flexibility in selecting a sampling algorithm.

During operation, multiple sensor reads per second are streamed from the FPGA to a readout PC, using the USB 3.0 port on the interface board. Alternate approaches using the FPGA to analyze the sensor reads and send processed data (ie SUTR slopes or Fowler averages) were also investigated, but were abandoned in favor of simpler firmware code after laboratory tests confirmed that the USB 3.0 bandwidth could handle the full data volume of streaming raw reads. Moving the data reduction to the FPGA may be appropriate for similar camera systems with more stringent constraints on computing power or communication bandwidth such as on space-based or suborbital platforms. For WINTER, the USB 3.0 output from each FPGA interface is converted to an OM4 multimode fiber, using an Icron\footnote{Icron/Maxim Integrated, San Jose, CA USA}USB 3.0 Spectra 3022 converter. The \textsim30 m fiber transmits the data from the instrument to a pair of processing PCs (1 PC per 3 sensors) in an electronics shed separated from the telescope dome (see Section \ref{sec: observatory}). These PCs process the raw sensor reads into FITS format images which are downlinked from the observatory in near-real-time.

The FPGA interfaces to the InGaAs sensors via a series of custom electronics boards designed by the WINTER team (see Ref. \citenum{winter_sensors} for further details). Each sensor has (a) a ``sensor board" which interfaces with the FLIR sensor and serves as a first stage preamplifier and motherboard for the additional boards, (b) an individual ``power board" to sequence the timing of applied voltages during power up/off, regulate the input voltages, and provide transient voltage suppression and overcurrent protection for the sensor in the event of static dissipation or lightning strikes to the observatory, and (c) an ``analog front end (AFE)" board with a second stage amplifier and analog to digital converters for output to the FPGA. Each sensor has its own readout electronics boards, forming six individual focal plane modules (FPMs).

\subsection{Thermal Management}
\label{subsec: thermal}

To reduce the dark current sufficiently to achieve sky-background-limited observations, the WINTER sensors are cooled with a two-stage TEC\footnote{Laird Thermal Systems, Durham, NC USA} mounted inside the vacuum-packaged sensor housing. Measurements with the DuPont Prototype showed that FLIR's InGaAs dark current falls below the Y-band sky at -40 \degC, but halves for each additional 7 degrees of cooling.\cite{Simcoe_2019, Sullivan_2013} Additional cooling below this transition point yields direct increases in sensitivity.  A proportional-integral-derivative (PID) loop running in software on the sensor PC communicates with the FPGA interface board to control the TEC current supplied by the power board on each FPM. The sensor packages cooled below room temperature to \textsim10 \degC to improve the cooling capacity of the TEC, while still keeping the delicate sensor above the 90th percentile dew point (\textsim 9 \degC) at Palomar Observatory.\footnote{Based on 5 years of Palomar Observatory 60 inch telescope weather station data from \url{http://bianca.palomar.caltech.edu/maintenance/weather/user_gen_file.tcl}} We expect the TEC to generate 10-12 W in operation, and a cold side temperature of \textsim50 \degC. The 10\degC base temperature is maintained with a Opti Temp\footnote{Opti Temp Inc., Traverse City, MI USA} OTC 1.0A chiller running a 30\% propylene glycol/distilled water cooling loop. To avoid coolant leaks within the instrument, a sintered copper/water heat pipe connects the sensor package to a liquid heat exchanger mounted outside the instrument enclosure (see Fig. \ref{figure:sensor}). 

%% file: Camera.tex
\section{Camera}
\label{sec: camera}
\subsection{Camera Optics}
\label{subsec: optics}
\begin{figure}[ht]
\centering
\includegraphics[width=0.75\linewidth]{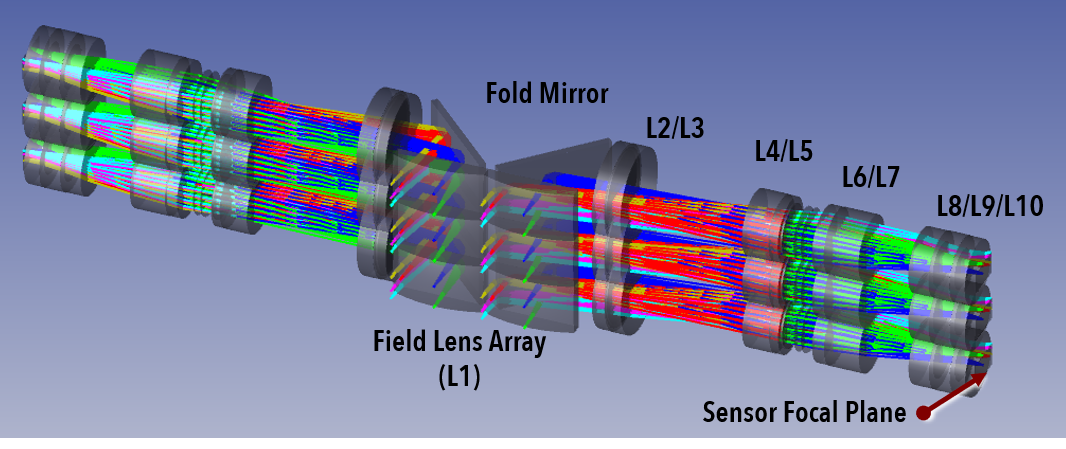}
\caption[Camera Optics Layout] {Ray trace from Zemax of the WINTER camera, showing the FLA near the telescope focal plane, the fold mirror which splits the instrument into two arms, and the individual lens arrays.}
\label{figure:optics}
\end{figure}

The driving requirement of the WINTER camera design is to maintain a high fill-factor field of view (\textgreater90\% over a \textsim1\deg x 1\deg FoV) despite not being able to directly abut the sensor elements. The design approach makes use of a novel ``fly's eye" approach built from a series of identical, replicated sensor channels. The layout is shown in Figure \ref{figure:optics}. In this replicated approach, the telescope focal plane is sliced into six individual channels, one per sensor, with a bonded array of plano-convex field lenses. Each individual lens has the same aspect ratio as the HD-format sensors, and the bonded array fills the full 1\deg FoV of the telescope. A right angle fold mirror splits the instrument into two separate arms, which improves access to the optical components during alignment, integration and testing (AIT), while also reducing the mechanical moment of the instrument by keeping the instrument center of mass closer to the telescope interface (see telescope description in Section \ref{subsec: telescope}). The field lens array (FLA) feeds six identical, replicated lens channels, each with their own sensor. Each channel has 10  lenses with spherical only surfaces (labeled L1 through L10 in the diagram), comprising a collimator section (L1 through L5) which forms a pupil near L5, and a camera section (L4 through L10). The 2:1 focal ratio of the collimator to camera sections reimage the F/6 telescope beam to F/3 at the sensor, providing a \textsim 1\arcsec/pixel plate scale. The degree of symmetry about the instrument pupil recalls opposing paired Petzval designs, with glasses selected to improve thermal stability. 

\subsubsection{Optimization}
A series of optimizations was carried out in Zemax\footnote{Zemax LLC, Kirkland, WA USA} based on the requirement flowdown from WINTER's science requirements. Key constraints in the design optimization were to achieve the 2:1 focal reduction, maintain an overall instrument length of less than \textsim1 m, and maintain sufficient space between telescope focus and the field lens vertex so that the instrument could be aligned optically separate from the telescope. To prevent cross-talk between channels marginal rays near the edge of the field lens must exit the lens either parallel, or aimed towards the optical axis. This also ensures that the pupil is smaller than the an individual field lens, preventing mechanical interference between stacked channels. 

\begin{figure}[ht]
\centering
\includegraphics[width=\linewidth]{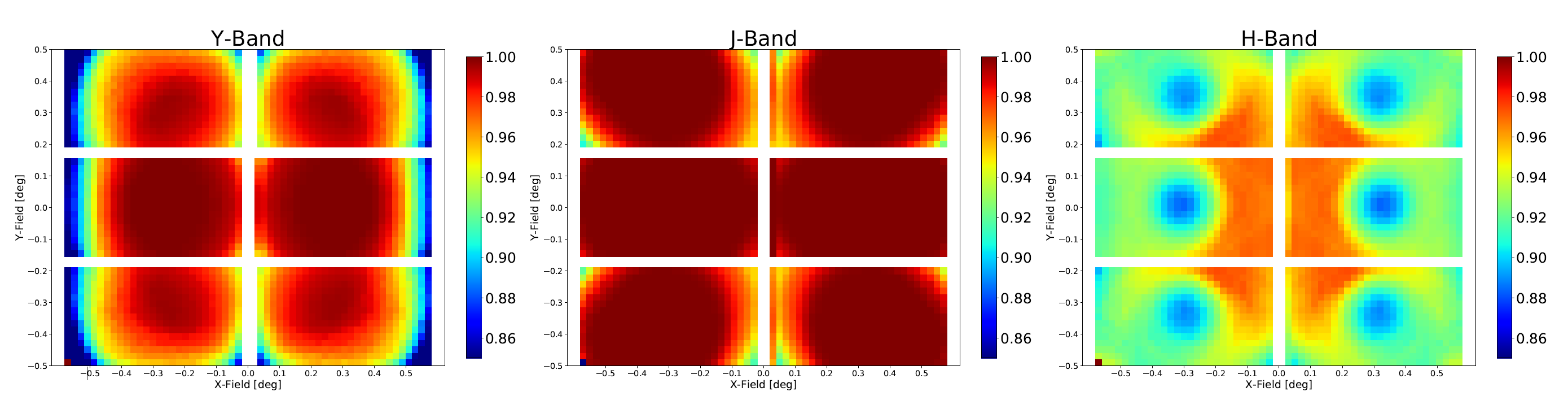}
\caption[Ensquared Energy] {Ensquared energy within 1 pixel for the CDR WINTER optical design over the full FoV for each of the three wavebands. Small changes in performance are expected based on a final reoptimization based on detailed index of refraction measurements of samples from each of the as-fabricated purchased glasses}.
\label{figure: image quality}
\end{figure}

Image quality requirements for the WINTER optical system derive from specific requirements on photometric mapping speed. This requirement definition used a model-based systems engineering approach and is described in detail in Ref. \citenum{winter_reqs}. 
The result of these optimizations achieves high and uniform image quality over the full FoV, as shown in Figure \ref{figure: image quality}, and serves as the reference optical design for the WINTER camera. After passing an internal critical design review (CDR), this design was used to purchase the glass blanks and begin the fabrication process. As described in the following section, one last round of optimization was carried out to rebalance the lens parameters based on the measured properties of the purchased glasses.

\subsubsection{Tolerancing}
\label{subsec: tolerancing}

A detailed study was carried out to establish tolerances on the optical glass selection, lens fabrication, optomechanical support structures, and AIT. The tolerance study was carried out using Zemax's Monte Carlo simulation capability. The studies were scripted to simulate different alignment procedures to select the optimal approach. This approach enforces the sequential nature of laboratory alignment and includes random errors in the optical compensation to more accurately represent the range of expected results. 

Based on the tolerance studies, an approach was selected which prioritized reducing the complexity of optical AIT at the cost of increased precision in the lens fabrication process. By maintaining strict fabrication tolerances, and obtaining precise measurements of the glass indices over the full WINTER passband, the mechanical alignment tolerances were loosened to the scale of typical machining tolerances (\textsim25-50 \um). This approach requires a single compensating element which is precisely adjusted independently for each channel. Based on the simulations, the horizontal and vertical (with respect to the optical axis) positioning of the L6/L7 bonded doublet were selected as the compensating degrees of freedom, requiring $\pm$5 \um positioning accuracy. The most sensitive aspect of the optical design is the index of refraction of the optical glasses. To reduce the effect of uncertainty in the glass parameters on the performance of the camera, all glass blanks (for each type of glass) were purchased from the same melt batch from the glass manufacturer\footnote{All glasses used in the WINTER design are from Ohara Corporation, Branchburg,
New Jersey USA}. Additionally, samples of each purchased glass type were sent to a third-party measurement shop\footnote{M$^3$ Measurement Solutions Inc., Escondido, CA USA} and measured at select optical test wavelengths and at regular intervals over the complete WINTER passband. These measurements reduced the uncertainty in the index of refraction from 0.5\% to 0.05\%. After recieving the measurements, a final round of optical optimization was carried out before finalizing the lens parameters.

\subsection{Optomechanics}
\label{subsec: optomech}

The close vertical packing of the individual channels within each arm requires that all lenses after the FLA in the optical path have truncations cut into the upper and lower edges. Based on the tolerance analysis, all lens groups except for the compensators, were cemented together into three-channel monoliths (the ``tiled lenses"). This is shown in Figure \ref{figure:optics}, where within each of the two arms of the instrument, all L2/3 groupings are bonded together, as are L4/L5 and L8/L9/L10. This approach is enabled by precision fabrication and alignment (\textless5\um placement errors) capabilities of the lens manufacturer, and was developed in collaboration with the selected vendor, Optimax Systems.\footnote{Optimax Systems Inc., Ontario, NY USA} Only the compensator lenses are independently mounted for each channel. The optomechanical implementation of the WINTER lens system is described in detail in Ref. \citenum{winter_optics}.

\subsection{Filters}
\label{subsec: filters}

\begin{wraptable}{R}{0.4\textwidth}
\caption{WINTER bandpass definitions.}
\begin{center}
\begin{tabular}{cccc}\\\toprule  
$\boldsymbol{\lambda}$ \textbf{(}$\boldsymbol{\upmu}$\textbf{m)} & \textbf{Center} & \textbf{Cut-On} & \textbf{Cut-Off}\\
\midrule \midrule
Y     & 1.02  & 0.97  & 1.07 \\
J     & 1.25  & 1.17  & 1.33 \\
Hs    & 1.60  & 1.49  & 1.68 \\
\bottomrule
\end{tabular}
\end{center}
\label{table: bandpass}
\end{wraptable} 

\begin{figure}[ht]
\centering
\includegraphics[width=\linewidth]{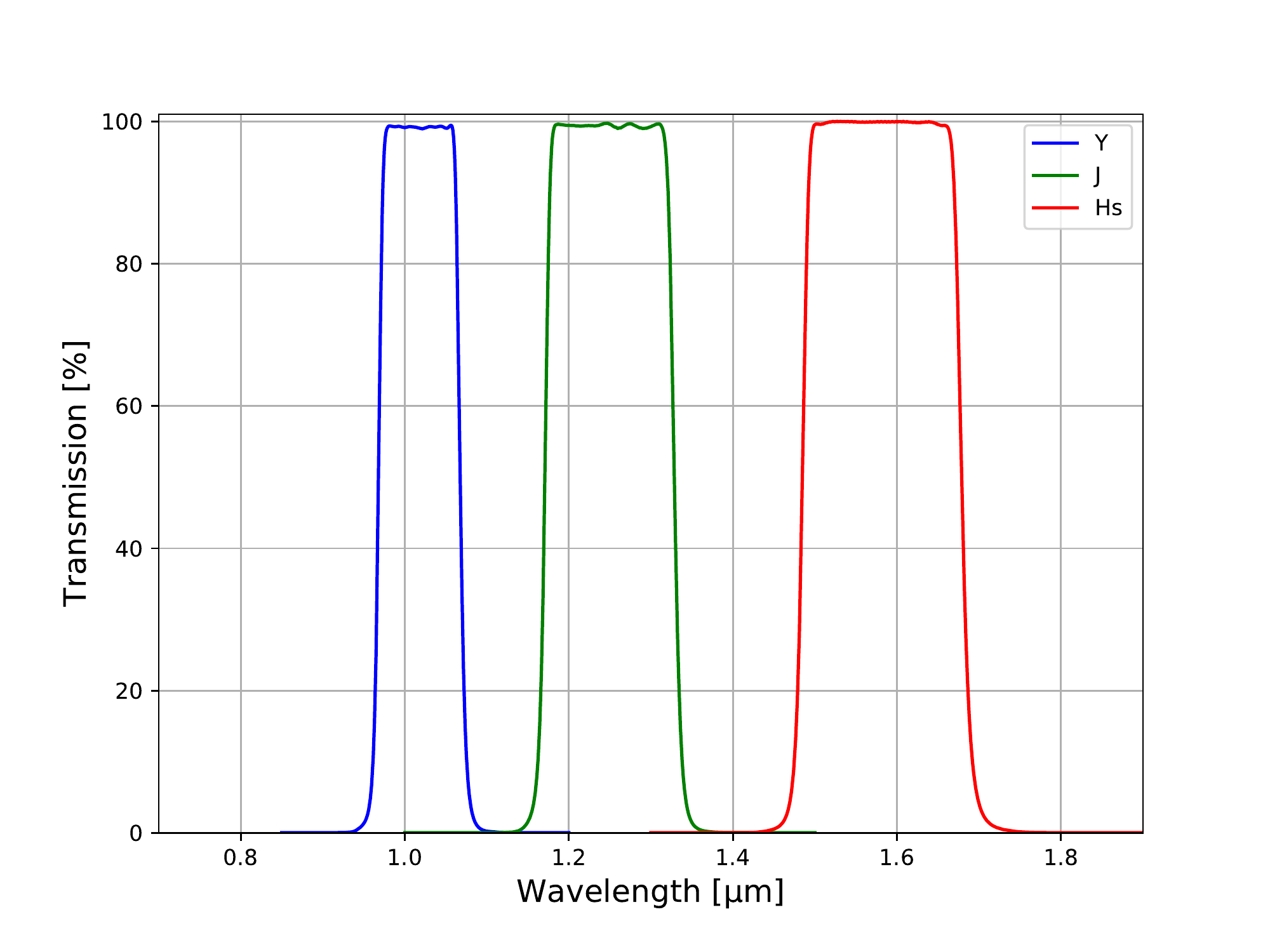}
\caption[Filter Bandpass] {Measured transmission for each of the three WINTER filters. Transmission data are measured with a parallel beam, and will deviate slightly (\textless0.5\%) in the WINTER F/6 beam. Data provided by the filter manufacturer, Asahi Spectra Inc.}
\label{figure: filters}
\end{figure}

The WINTER filters are based on the canonical Mauna Kea Observatory (MKO) filter set\cite{tokunaga_mko}, with a modified shortened H-band (Hs-band) with a long-wave cutoff tuned to the 1.7 \um InGaAs bandgap cutoff. A single 3-position filter tray sits 50 mm in front of the FLA in the converging, telecentric, F/6 beam of the telescope. While the $\pm$4.8\deg angle of incidence (AOI) variation with the F/6 beam somewhat reduces the slope of the band edges, placing the filter in the converging beam ensures spectral uniformity across the FoV, simplifying photometric calibration.  The custom WINTER bandpass filters were built by Asahi Spectra Co.\footnote{Asahi Spectra Co., Ltd. Tokyo, Japan} on 10 mm thick fused silica, and achieve \textgreater 99\% throughput in band with band-edge slopes ($\lambda_{90\% \: \mathrm{Trans}} - \lambda_{10\% \: \mathrm{Trans}})/\lambda_{10\% \: \mathrm{Trans}}$ \textsim1\%, and block out-of-band light to \textless0.01\% (OD4) over an extended range from 0.7 to 1.9 \um to prevent blue leaks at shorter wavelengths where the InGaAs detector is still sensitive. The bandpass specifications are listed in Table \ref{table: bandpass}, and the measured spectra are shown in Fig. \ref{figure: filters}.

%% file: Observatory.tex
\section{Observatory}
\label{sec: observatory}

\subsection{Telescope}
\label{subsec: telescope}
\begin{figure}[ht]
\centering
\includegraphics[width=\linewidth]{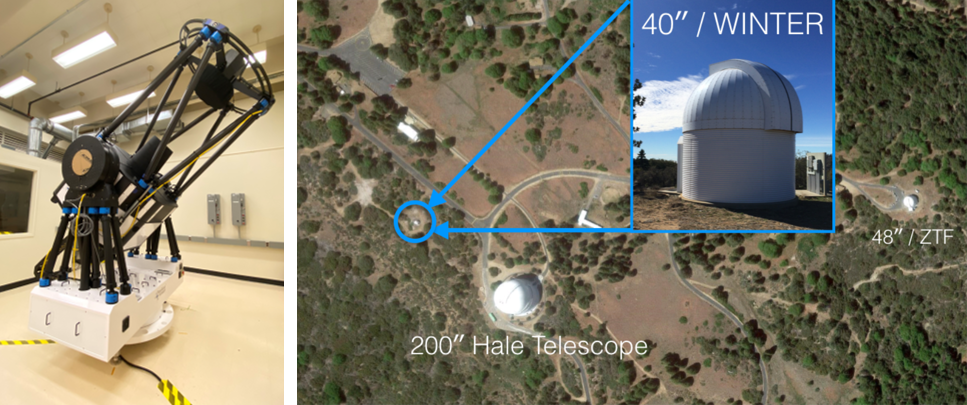}
\caption[Telescope and Observatory] {\textbf{Left:} The WINTER 1 m PlaneWave PW1000 telescope during laboratory testing at MIT. \textbf{Right:} Site layout of the WINTER site at Palomar Observatories indicating the location of the WINTER enclosure, and photo of the telescope pad and dome.}
\label{figure:telescope}
\end{figure}

WINTER uses a COTS 1 m corrected Dall-Kirkham (CDK) telescope with fast direct-drive altitude/azimuth pointing motors, and a instrument mount which compensates for sky rotation. The telescope has all fused silica optics, including two sets of three corrector lenses mounted in lens barrels within each of the two Nasmyth platforms. A flat, tertiary (M3) mirror can steer the beam towards either Nasmyth port. PlaneWave Instruments\footnote{PlaneWave Instruments, Adrian, MI USA} made several modifications to their base PW1000 1-Meter Observatory System for WINTER, including optimizing the mirror antireflective (AR) coatings for infrared operation and developing a M2 focusing system. The Nasmyth port where the WINTER instrument is mounted has corrector lenses with IR-optimized AR coatings, while the opposite port is has optical-optimized coatings to allow for a future visible-light instrument to be mounted there. The main telescope components and instrument mount are shown in Fig. \ref{figure:winter_overview}, and a photograph of the telescope during laboratory testing is shown in Fig. \ref{figure:telescope}.

The telescope is controlled by a small embedded Windows PC\footnote{ARK-1550-9551, Advantech Co. Ltd - North America, Milpitas, CA USA} which runs a \texttt{http} command server from PlaneWave. This server takes pointing and operating commands from the main WINTER observatory control software described in Section \ref{subsec: controls}.

\subsection{Site}
\label{subsec: site}
WINTER will operate at Palomar Observatory at an altitude of \textsim 1700 m. The instrument will be installed in an existing 5 m dome near the 200 inch Hale Telescope operated by the Caltech Optical Observatories (COO). This dome has been updated with new motor drives, a custom pier, a weather station, and a telemetry and robotic operating system similar to those currently developed by COO for the 60 inch robotic telescope, the nearby Zwicky Transient Facility (ZTF)\cite{ztf_overview_Bellm_2018}, and PGIR\cite{Gattini_De_2020}. The dome communications and operating protocols derived from WINTER's observational goals are detailed in Ref. \citenum{winter_reqs}. A temperature-controlled shed a few meters from the dome houses all of the control electronics for the WINTER instrument, including the subsystem control PCs, instrument power supplies, and chiller for the sensor liquid coolant loop.

\subsection{Control System}
\label{subsec: controls}

WINTER's robotic operating system is controlled by a custom control program. This program, the WINTER Supervisory Program (WSP), is written in Python 3.7 and uses \texttt{PyQt5}\footnote{Riverbank Computing, \url{https://www.riverbankcomputing.com/software/pyqt/}}, a set of Python bindings to the Qt C++ application framework\footnote{Qt Group, \url{https://www.qt.io/}}, enabling multithreaded operation. A series of cron jobs on the observatory control PC run schedule production code to generate a nightly schedule (see Section \ref{subsec: scheduling}), and initiate the WSP. Elements of the control software, including the weather decision-making and communications with the telescope are adapted from the MINERVA project,  telescopes.\cite{MINERVA_Swift_2015,MINERVA_Wilson_2019}, which share heritage with the Robo-AO system which operates the Palomar 60 inch observatory\cite{robo-ao_Riddle_2012}. While the WSP is designed to make appropriate decisions about when to open the dome and initiate observations, the robotic control can be overridden by the on-site operator at the nearby 200 inch Hale Telescope who will close the WINTER dome in the event of inclement weather or other undetected problems.

The WSP runs a series of threads to (1) load the current scheduled observation (2) listen for target of opportunity (ToO) events (3) listen for commands from the terminal and externally over TCP/IP, (4) execute these commands from a priority queue and dispatch commands to the sensor and telescope PCs, (5) log housekeeping data about the telescope and instrument state to a telemetry database, and (6) log all observations to a observation database. The telemetry database is stored in dirfile binary format using GetData \texttt{PyGetData}\footnote{The GetData Project, \url{http://getdata.sourceforge.net/}} Python bindings. This binary database format can be used with a number of software tools developed at the University of Toronto for stratospheric balloon programs\cite{toronto_software_romualdez}, including a realtime plotting tool, Kst\footnote{https://kst-plot.kde.org/}, and \texttt{owl},\footnote{https://github.com/BlastTNG/flight/tree/master/owl} a modular dashboard program. The observation log is a SQlite database which can easily be queried, or even completely copied and transferred off the control computer for analysis. It is also used by the scheduler program to further optimize observing plans for future nights, and recover from interrupted schedules. The structure of the database is defined in a separate configuration file to allow for easy modification of the database schema, without having to modify the software responsible for recording observations.

%% file: Observing.tex
\section{Observing Strategy}
\label{sec: observing}
WINTER's preliminary observing plan interweaves two systematic surveys: a \textbf{Shallow-Wide}  survey in WINTER's deepest filter (J), and a \textbf{Deep-Fast} survey in all three bands. The Shallow-Wide survey will cover 4050 square degrees at a 9-day cadence, stepping fields every four epochs to cover the full northern sky in the first year of operations to $\mathrm{J_{AB}} = $ 19.2 (single-visit) or 20.5 (coadded). Completing this survey in the first year is key to building a sufficiently deep sky map to enable transient detections over WINTER's full volume by the time LIGO-Virgo O4 begins. The Deep-Fast survey in Y+J+Hs observes 450 \degsq at a 3-day cadence, rotating to new fields after twenty epochs.

\subsection{Scheduling}
\label{subsec: scheduling}
To determine an optimized observing for multiple simultaneous observing programs, WINTER uses a customized scheduler that builds on the extensive functionality of the ZTF scheduler.\cite{ztf_scheduler_Bellm_2019} The ZTF scheduler balances many different observing programs with unique cadences while also maximizing data quality by selecting fields through volumetric weighting. In this scheme, the observable sky is split into discrete fields matching the instrument FoV. At any time, the most desirable field to observe is that which probes the greatest limiting volume for any given exposure. The limiting volume for an exposure is related to the distance  at which a source of absolute magnitude M will be detected. This is constrained by seeing, sky brightness, and instrument noise, which all factor into the limiting magnitude (m$_{\mathrm{lim}}$) for an exposure.

The ZTF scheduler offers several modes for simulations and on-sky observing: (1) Queue
observing ingests a predefined list of fields and steps through each field sequentially, (2) ``Greedy" observing continuously selects the best target (based on the volumetric weighting scheme) for a given time and recalculating before each target, (3) Optimized observing uses the Gurobi\cite{gurobi} linear optimizer to solve a travelling salesperson problem to optimize each night of observing. All three modes are used in WINTER’s observing program: queue observing steps through observations for defined targets of opportunity (ToO), such as LIGO alerts, optimized observing balances science surveys with varying cadences, and greedy observing fills in reference images of the sky between science surveys and ToOs.

This scheduling code is used for simulating entire observing programs, as well as producing a nightly schedule. To develop and optimize observing program design, simulations of several years of observing are carried out using historical weather data from Palomar Observatory to realistically estimate the typical observing conditions throughout the year. For actual observing, an optimized schedule is created each night which incorporates the recorded data in the observation database. 

For catching GW detection alerts and scheduling ToO observations, WINTER uses the GROWTH ToO Marshal\cite{GROWTH_Marshal_Coughlin_2019,GROWTH_Marshal_Kasliwal_2019}. The GROWTH Marshal web portal ingests GW alerts distributed via the LIGO-Virgo GCN system\footnote{LIGO Scientific Collaboration Public Alerts, \url{https://emfollow.docs.ligo.org/userguide/index.html}}, and uses the GW EM counterpart search package \texttt{gwemopt}\cite{gwemopt_Coughlin_2018} as a backend to help produce optimized observing schedules for WINTER and other kilonova discovery engines such as ZTF, PGIR, and DECam. The ToO marshal is also integrated into WINTER's analysis pipeline (see Section \ref{subsec: image processing}) to publish detections of transient candidates. Because the GROWTH Marshal software is tuned towards visible wavelength followup observations which peak in the first few days after a BNS merger, some modifications to the approach will be necessary to tune the observations for longer-lasting IR followup. Depending on the frequency of ToO events, and the performance of the upgraded LIGO O4 survey, it may be necessary to add additional features to the WINTER scheduling architecture, including a galaxy-weighted wide-field search mode which weighs individual fields based on published galaxy catalogs while still taking advantage of WINTER's ability to cover the full LIGO positional error contour, and the ability to interleave multiple ToO schedules to handle overlapping events. This ongoing development effort will continue through WINTER's commissioning and first year of observations.

The ToO schedules are written in the same ZTF-style database format as the nightly schedules and are loaded the same way into the WSP software. As soon as all possible observations in the ToO schedule are completed, the observatory will switch over automatically to the nightly schedule.

\begin{figure}[ht]
\centering
\includegraphics[width=\linewidth]{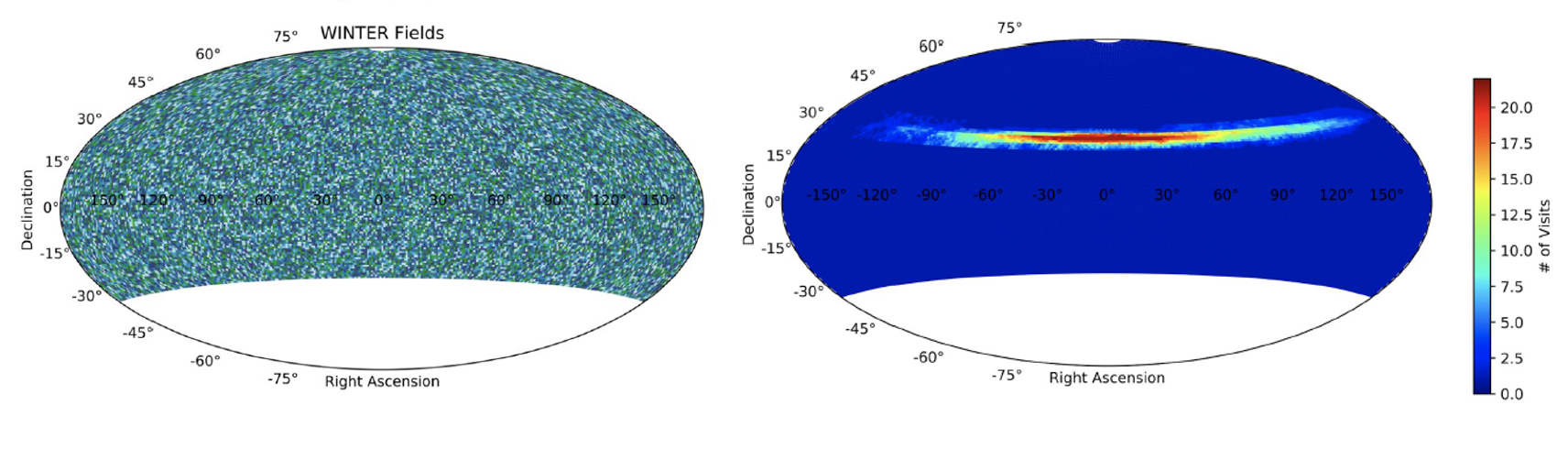}
\caption[Scheduler Results] {\textbf{Left:} WINTER's available night sky split into \textsim35,000 individually numbered 1\deg x 1\deg fields, used for schedule optimization and image cataloguing. \textbf{Right:} A sample hit map from a optimization run simulating a the first year of WINTER observations. The blue background represents the creation of a deep J-band series of reference images for the northern sky. The high-visit-density stripe represents a deep-fast observing program optimized to select patches of sky that can be revisited every three nights. Historical weather data from Palomar Observatory is used to make reasonable estimates of favorable observing conditions.}
\label{figure:scheduler}
\end{figure}%

\subsection{Image Processing}
\label{subsec: image processing}

The WINTER data processing approach draws on the heritage of the mature transient detection and analysis pipelines  developed  at  Caltech  for  the  successful ZTF \cite{ztf_data_pipeline_Masci_2018} and PGIR \cite{Gattini_De_2020}) programs. Similar to PGIR, each WINTER observation will be a series of dithered exposures to facilitate longer exposure times on the bright sky background. Images acquired in each dithered sequence will be stacked using the Drizzle \cite{Fruchter2002} algorithm. The Drizzle algorithm will enable reconstruction of the point spread function (PSF) in images where the PSF is undersampled. Astrometric and photometric calibration will be carried out using the 2MASS point source catalog \cite{Cutri2003} for the J and H filters and the Pan-STARRS catalog \cite{Pan-STARRS_Flewelling2020} for the Y filter. For each field, data taken from at least five repeat visits evenly spaced during the first year of observations will be stacked to build deep ($\approx 21$ mag) reference images. All subsequent ``science" images will be fed to the image differencing pipeline, where they will be subtracted from the corresponding reference image using the ZOGY algorithm \cite{Zackay2016}. The subtracted image will be then fed to the Astromatic package SExtractor \cite{Bertin1996} to detect candidate transient sources. A machine-learning based classifier will be used to help automatically distinguish between real astrophysical transient sources and image subtraction artifacts. The candidates that are flagged as real transients will be uploaded to a web-portal for human vetting. Candidate vetting and follow-up by various telescopes will be coordinated by a web portal similar to the GROWTH Marshal\cite{GROWTH_Marshal_Kasliwal_2019,GROWTH_Marshal_Coughlin_2019}. A brand new Fritz system, the next generation of the GROWTH Marshal, is currently being developed as an open source project that integrates and extends two projects: \texttt{Kowalski}\cite{Duev_2019} and SkyPortal\cite{skyportal}.

%% file: Status.tex
\section{Current Status}
\label{sec: status}
The WINTER instrument development effort is ongoing, with a planned commissioning date of summer 2021. The telescope and observatory control system is currently undergoing laboratory tests at MIT, and be installed at the site once infrastructure improvements at the Palomar Observatory site are completed in early 2021. We are currently stress-testing the robotic system which takes in real-time weather and site data from the Palomar Observatory and the WINTER dome from dedicated telemetry servers. The  WINTER camera lenses are currently in fabrication, and we expect the camera optics to be integrated during late spring 2021. The optomechanical mounting approach for both the three-channel bonded lenses and the individually-mounted compensator lenses has been demonstrated through detailed prototyping and characterization, as described in Ref. \citenum{winter_optics}. The WINTER sensors are currently being fabricated, and the readout electronics and firmware have been prototyped and are currently being tested (see Ref. \citenum{winter_sensors}) ahead of upcoming systems-level tests of the focal plane modules with engineering-grade sensors. 

%% file: Conclusion.tex
\section{Conclusion}
\label{sec: conclusion}
WINTER will be a powerful new tool for researching the transient sky in the infrared, a relatively unexplored regime in which a host of astrophysical phenomena are most visible, including the thermal emission from \rprocess element synthesis in the ejecta of neutron star mergers. Observing in the Y, J, and Hs bands near the expected emission peak of BNS kilonovae, WINTER will robotically execute follow-up observations of GW events detected by the LIGO-Virgo instruments. WINTER is the first seeing-limited instrument dedicated to systematic infrared time-domain searches, and will perform multiple simultaneous observing programs to identify IR transients. WINTER's planned commissioning beginning in summer 2021 will allow sufficient time to build a full map of the available sky in J-band before the LIGO-Virgo O4 observing run, enabling WINTER's full kilonova search capability in its first year of science operations. WINTER is part of a new generation of IR time-domain instruments, including DREAMS at Siding Spring Observatory in New South Wales, Australia\cite{DREAMS_Soon_2018}, PRIME\cite{PRIME} on the South African Astronomical Observatory, and proposed Antarctic instruments like Cryoscope\cite{cryoscope_smith_2020} which will open new windows into the transient sky.